%Paper: hep-th/9207093
%From: TJALLEN@wishep.physics.wisc.edu
%Date: Tue, 28 Jul 92 13:55 CDT
%Date (revised): Tue, 28 Jul 92 14:36 CDT

%%%%%%%%%%%%%%%%%%%%%%%%%%%%%%%%%%%%%%%%%%%%%%%%%%%%%%%%%%%%%%%%%%%%%%%%%
% % % % % % % % % % % % % % % % % % % % % % % % % % % % % % % % % % % %
%%%   This is PHYZZX macro package.   % % % % % % % % % % % % % % % % %
%% % % % % % % % % % % % % % % % % % % % % % % % % % % % % % % % % % % %
%%%  This version of PHYZZX should be used with Version >1.0 of TEX % %
%% % % % % % % % % % % % % % % % % % % % % % % % % % % % % % % % % % % %
%%%   To preload both PLAIN and PHYZZX, begin your file with    % % % %
%%%  a line "%macropackage=phyzzx" instead of "\input phyzzx".  % % % %
%% % % % % % % % % % % % % % % % % % % % % % % % % % % % % % % % % % % %
%%%%%%%%%%%%%%%%%%%%%%%%%%%%%%%%%%%%%%%%%%%%%%%%%%%%%%%%%%%%%%%%%%%%%%%%
%%%%%%%  Created by Vadim Kaplunovsky in June 1984.   %%%%%%%%%%%%%%%%%%
% % % % % % % % % % % % % % % % % % % % % % % % % % % % % % % % % % % %
%%%%%%%%%%%%  Latest update/debug: April 27, 1988   %%%%%%%%%%%%%%%%%%%%
%%%%%%%%%%%%%%%%%%%%%%%%%%%%%%%%%%%%%%%%%%%%%%%%%%%%%%%%%%%%%%%%%%%%%%%%
%
\expandafter\ifx\csname phyzzx\endcsname\relax
 \message{It is better to use PHYZZX format than to
          \string\input\space PHYZZX}\else
 \wlog{PHYZZX macros are already loaded and are not
          \string\input\space again}%
 \endinput \fi
\catcode`\@=11 % This allows us to modify PLAIN macros.
\let\rel@x=\relax
\let\n@expand=\relax
\def\pr@tect{\let\n@expand=\noexpand}
\let\protect=\pr@tect
\let\gl@bal=\global
%
%%%%%%%%%%%%%%%%%%%%%%%%%%%%%%%%%%%%%%%%%%%%%%%%%%%%%%%%%%%%%%%%%%%%%%%%
%
% First, I define fonts and basic spacing parameters
%
\newfam\cpfam
\newdimen\b@gheight             \b@gheight=12pt
\newcount\f@ntkey               \f@ntkey=0
\def\f@m{\afterassignment\samef@nt\f@ntkey=}
\def\samef@nt{\fam=\f@ntkey \the\textfont\f@ntkey\rel@x}
\def\setstr@t{\setbox\strutbox=\hbox{\vrule height 0.85\b@gheight
                                depth 0.35\b@gheight width\z@ }}
%
%
% PHYZZX fonts are kept in this separate file
% in order to facilitate font substitution.
%
% This file should be called PHYZZX.FONTS on sites using CM fonts
% and PHYZZX.CMFONTS on other sites.
%
%
% fourteenpoint font substitutions made for the talaris printer
%
%%%%%%%%%%%%%%%%%%%%%%%%%%%%%%%%%%%%%%%%%%%%%%%%%%%%%
\newfam\ssfam   % Define a San Serif font family
%
%%%%%%%%%%%%% Computer Modern Roman %%%%%%%%%%%%%%%%%%%%%%%%%%%%%%%%
\font\seventeenrm =cmr17
%\font\fourteenrm  =cmr12 scaled\magstep1
\font\fourteenrm  =cmr10 scaled\magstep2 % was cmr12 scaled\magstep1
\font\twelverm    =cmr12
\font\ninerm      =cmr9
\font\sixrm       =cmr6
\font\fiverm      =cmr5
%
%%%%%%%%%%%%%%%%% San Serif %%%%%%%%%%%%%%%%%%%%%%%%%%%%%%%%%%%%%%%%%%%

\font\fourteenss  =cmss10 scaled\magstep2
\font\twelvess    =cmss10 scaled\magstep1
\font\tenss       =cmss10
\font\niness      =cmss9
\font\eightss     =cmss8
%
%%%%%%%%%%%%%%%%% San Serif Bold %%%%%%%%%%%%%%%%%%%%%%%%%%%%%%%%%%%%%%%

%
%
%%%%%%%%%%%%%%%%% San Serif Italic %%%%%%%%%%%%%%%%%%%%%%%%%%%%%%%%%%%%%%

%
%%%%%%%%%%%%% Computer Modern Bold  %%%%%%%%%%%%%%%%%%%%%%%%%%%%%%%%%%%%

%\font\fourteenbf  =cmbx12 scaled\magstep1
\font\fourteenbf  =cmbx10 scaled\magstep2 % was cmbx12 scaled\magstep1
\font\twelvebf    =cmbx12
\font\ninebf      =cmbx9
\font\sixbf       =cmbx6
%
%%%%%%%%%%%%%%%%% Computer Modern Math Italic %%%%%%%%%%%%%%%%%%%%%%%%%%
\font\seventeeni  =cmmi12 scaled\magstep2    \skewchar\seventeeni='177
%\font\fourteeni   =cmmi12 scaled\magstep1     \skewchar\fourteeni='177
\font\fourteeni   =cmmi10 scaled\magstep2     \skewchar\fourteeni='177
\font\twelvei     =cmmi12                       \skewchar\twelvei='177
\font\ninei       =cmmi9                          \skewchar\ninei='177
\font\sixi        =cmmi6                           \skewchar\sixi='177
%
%%%%%%%%%%%%%%%%%%% Math Symbols %%%%%%%%%%%%%%%%%%%%%%%%%%%%%%%%%%%%%%%
\font\seventeensy =cmsy10 scaled\magstep3    \skewchar\seventeensy='60
\font\fourteensy  =cmsy10 scaled\magstep2     \skewchar\fourteensy='60
\font\twelvesy    =cmsy10 scaled\magstep1       \skewchar\twelvesy='60
\font\ninesy      =cmsy9                          \skewchar\ninesy='60
\font\sixsy       =cmsy6                           \skewchar\sixsy='60
%
%%%%%%%%%%%%%%%%%%% Extended Math Symbols %%%%%%%%%%%%%%%%%%%%%%%%%%%%%%%

\font\fourteenex  =cmex10 scaled\magstep2
\font\twelveex    =cmex10 scaled\magstep1
%
%%%%%%%%%%%%%%%%% Computer Modern Slanted %%%%%%%%%%%%%%%%%%%%%%%%%%%%%%%

%\font\seventeensl =cmsl12 scaled\magstep2
%\font\fourteensl  =cmsl12 scaled\magstep1
\font\fourteensl  =cmsl10 scaled\magstep2
\font\twelvesl    =cmsl12
\font\ninesl      =cmsl9
%
%%%%%%%%%%%%%%%%% Computer Modern Italic  %%%%%%%%%%%%%%%%%%%%%%%%%%%%%%%

%\font\seventeenit =cmti12 scaled\magstep2
%\font\fourteenit  =cmti12 scaled\magstep1
\font\fourteenit  =cmti10 scaled\magstep2 % for talaris
\font\twelveit    =cmti12
\font\nineit      =cmti9
%
%%%%%%%%%%%%%%%%%%% Typewriter Fonts %%%%%%%%%%%%%%%%%%%%%%%%%%%%%%%%%%%%
%\font\fourteentt  =cmtt12 scaled\magstep1
\font\fourteentt   =cmtt10 scaled\magstep2
\font\twelvett     =cmtt12
\font\tentt        =cmtt10
%
%%%%%%%%%%%%%%%%%% Small Capitals %%%%%%%%%%%%%%%%%%%%%%%%%%%%%%%%%%%%%%%
\font\fourteencp   =cmcsc10 scaled\magstep2
\font\twelvecp     =cmcsc10 scaled\magstep1
\font\tencp        =cmcsc10
%
%
%%%%%%%%%%%%%%%%%%%%%%%%%%%%%%%%%%%%%%%%%%%%%%%%%%%%%%%%%%
%
\def\fourteenf@nts{\relax
    \textfont0=\fourteenrm          \scriptfont0=\tenrm
      \scriptscriptfont0=\sevenrm
    \textfont1=\fourteeni           \scriptfont1=\teni
      \scriptscriptfont1=\seveni
    \textfont2=\fourteensy          \scriptfont2=\tensy
      \scriptscriptfont2=\sevensy
    \textfont3=\fourteenex          \scriptfont3=\twelveex
      \scriptscriptfont3=\tenex
    \textfont\itfam=\fourteenit     \scriptfont\itfam=\tenit
    \textfont\slfam=\fourteensl     \scriptfont\slfam=\tensl
    \textfont\bffam=\fourteenbf     \scriptfont\bffam=\tenbf
      \scriptscriptfont\bffam=\sevenbf
    \textfont\ttfam=\fourteentt
    \textfont\cpfam=\fourteencp
    \textfont\ssfam=\fourteenss     \scriptfont\ssfam=\tenss
        \scriptscriptfont\ssfam=\sevenrm }
\def\twelvef@nts{\relax
    \textfont0=\twelverm          \scriptfont0=\ninerm
      \scriptscriptfont0=\sixrm
    \textfont1=\twelvei           \scriptfont1=\ninei
      \scriptscriptfont1=\sixi
    \textfont2=\twelvesy           \scriptfont2=\ninesy
      \scriptscriptfont2=\sixsy
    \textfont3=\twelveex          \scriptfont3=\tenex
      \scriptscriptfont3=\tenex
    \textfont\itfam=\twelveit     \scriptfont\itfam=\nineit
    \textfont\slfam=\twelvesl     \scriptfont\slfam=\ninesl
    \textfont\bffam=\twelvebf     \scriptfont\bffam=\ninebf
      \scriptscriptfont\bffam=\sixbf
    \textfont\ttfam=\twelvett
    \textfont\cpfam=\twelvecp    \scriptfont\cpfam=\tencp
    \textfont\ssfam=\twelvess    \scriptfont\ssfam=\niness
      \scriptscriptfont\ssfam=\sixrm }
\def\tenf@nts{\relax
    \textfont0=\tenrm          \scriptfont0=\sevenrm
      \scriptscriptfont0=\fiverm
    \textfont1=\teni           \scriptfont1=\seveni
      \scriptscriptfont1=\fivei
    \textfont2=\tensy          \scriptfont2=\sevensy
      \scriptscriptfont2=\fivesy
    \textfont3=\tenex          \scriptfont3=\tenex
      \scriptscriptfont3=\tenex
    \textfont\itfam=\tenit     \scriptfont\itfam=\seveni  % no \sevenit
    \textfont\slfam=\tensl     \scriptfont\slfam=\sevenrm % no \sevensl
    \textfont\bffam=\tenbf     \scriptfont\bffam=\sevenbf
      \scriptscriptfont\bffam=\fivebf
    \textfont\ttfam=\tentt
    \textfont\cpfam=\tencp
    \textfont\ssfam=\tenss      \scriptfont\ssfam=\eightss
      \scriptscriptfont\ssfam=\fiverm }
\def\ss{\n@expand\f@m\ssfam}
%

% Actual font definitions are kept in a separate file
% to facilitate font substitution.
%
\def\rm{\n@expand\f@m0 }
\def\mit{\n@expand\f@m1 }         
\def\cal{\n@expand\f@m2 }
\def\it{\n@expand\f@m\itfam}
\def\sl{\n@expand\f@m\slfam}
\def\bf{\n@expand\f@m\bffam}
\def\tt{\n@expand\f@m\ttfam}
\def\caps{\n@expand\f@m\cpfam}    
\def\em@{\rel@x\ifnum\f@ntkey=0 \it \else
        \ifnum\f@ntkey=\bffam \it \else \rm \fi \fi }
\def\em{\n@expand\em@}
\def\fourteenpoint{\fourteenf@nts \samef@nt \b@gheight=14pt \setstr@t }
\def\twelvepoint{\twelvef@nts \samef@nt \b@gheight=12pt \setstr@t }
\def\tenpoint{\tenf@nts \samef@nt \b@gheight=10pt \setstr@t }
\normalbaselineskip = 20pt plus 0.2pt minus 0.1pt
\normallineskip = 1.5pt plus 0.1pt minus 0.1pt
\normallineskiplimit = 1.5pt
\newskip\normaldisplayskip
\normaldisplayskip = 20pt plus 5pt minus 10pt
\newskip\normaldispshortskip
\normaldispshortskip = 6pt plus 5pt
\newskip\normalparskip
\normalparskip = 6pt plus 2pt minus 1pt
\newskip\skipregister
\skipregister = 5pt plus 2pt minus 1.5pt
\newif\ifsingl@
\newif\ifdoubl@
\newif\iftwelv@  \twelv@true
\def\singlespace{\singl@true\doubl@false\spaces@t}
\def\doublespace{\singl@false\doubl@true\spaces@t}
\def\normalspace{\singl@false\doubl@false\spaces@t}
\def\Tenpoint{\tenpoint\twelv@false\spaces@t}
\def\Twelvepoint{\twelvepoint\twelv@true\spaces@t}
\def\spaces@t{\rel@x
      \iftwelv@ \ifsingl@\subspaces@t3:4;\else\subspaces@t1:1;\fi
       \else \ifsingl@\subspaces@t3:5;\else\subspaces@t4:5;\fi \fi
      \ifdoubl@ \multiply\baselineskip by 5
         \divide\baselineskip by 4 \fi }
\def\subspaces@t#1:#2;{\rel@x
      \baselineskip = \normalbaselineskip
      \multiply\baselineskip by #1 \divide\baselineskip by #2
      \lineskip = \normallineskip
      \multiply\lineskip by #1 \divide\lineskip by #2
      \lineskiplimit = \normallineskiplimit
      \multiply\lineskiplimit by #1 \divide\lineskiplimit by #2
      \parskip = \normalparskip
      \multiply\parskip by #1 \divide\parskip by #2
      \abovedisplayskip = \normaldisplayskip
      \multiply\abovedisplayskip by #1 \divide\abovedisplayskip by #2
      \belowdisplayskip = \abovedisplayskip
      \abovedisplayshortskip = \normaldispshortskip
      \multiply\abovedisplayshortskip by #1
        \divide\abovedisplayshortskip by #2
      \belowdisplayshortskip = \abovedisplayshortskip
      \advance\belowdisplayshortskip by \belowdisplayskip
      \divide\belowdisplayshortskip by 2
      \smallskipamount = \skipregister
      \multiply\smallskipamount by #1 \divide\smallskipamount by #2
      \medskipamount = \smallskipamount \multiply\medskipamount by 2
      \bigskipamount = \smallskipamount \multiply\bigskipamount by 4 }
\def\normalbaselines{ \baselineskip=\normalbaselineskip
   \lineskip=\normallineskip \lineskiplimit=\normallineskip
   \iftwelv@\else \multiply\baselineskip by 4 \divide\baselineskip by 5
     \multiply\lineskiplimit by 4 \divide\lineskiplimit by 5
     \multiply\lineskip by 4 \divide\lineskip by 5 \fi }
\Twelvepoint  % That's the default
\interlinepenalty=50
\interfootnotelinepenalty=5000
\predisplaypenalty=9000
\postdisplaypenalty=500
\hfuzz=1pt
\vfuzz=0.2pt
\newdimen\HOFFSET  \HOFFSET=0pt
\newdimen\VOFFSET  \VOFFSET=0pt
\newdimen\HSWING   \HSWING=0pt
\dimen\footins=8in
%
%%%%%%%%%%%%%%%%%%%%%%%%%%%%%%%%%%%%%%%%%%%%%%%%%%%%%%%%%%%%%%%%%%%%%%%%
%
%   Next, I define output routines, footnotes & related stuff.
%
\newskip\pagebottomfiller
\pagebottomfiller=\z@ plus \z@ minus \z@
\def\pagecontents{
   \ifvoid\topins\else\unvbox\topins\vskip\skip\topins\fi
   \dimen@ = \dp255 \unvbox255
   \vskip\pagebottomfiller
   \ifvoid\footins\else\vskip\skip\footins\footrule\unvbox\footins\fi
   \ifr@ggedbottom \kern-\dimen@ \vfil \fi }
\def\makeheadline{\vbox to 0pt{ \skip@=\topskip
      \advance\skip@ by -12pt \advance\skip@ by -2\normalbaselineskip
      \vskip\skip@ \line{\vbox to 12pt{}\the\headline} \vss
      }\nointerlineskip}
\def\makefootline{\baselineskip = 1.5\normalbaselineskip
                 \line{\the\footline}}
\newif\iffrontpage
\newif\ifp@genum
\def\nopagenumbers{\p@genumfalse}
\def\pagenumbers{\p@genumtrue}
\pagenumbers
\newtoks\paperheadline
\newtoks\paperfootline
\newtoks\letterheadline
\newtoks\letterfootline
\newtoks\letterinfo
\newtoks\date
\paperheadline={\hfil}
\paperfootline={\hss\iffrontpage\else\ifp@genum\tenrm\folio\hss\fi\fi}
\letterheadline{\iffrontpage \hfil \else
    \rm \ifp@genum page~~\folio\fi \hfil\the\date \fi}
\letterfootline={\iffrontpage\the\letterinfo\else\hfil\fi}
\letterinfo={\hfil}
\def\monthname{\rel@x\ifcase\month 0/\or January\or February\or
   March\or April\or May\or June\or July\or August\or September\or
   October\or November\or December\else\number\month/\fi}
\def\today{\monthname~\number\day, \number\year}
\date={\today}
\headline=\paperheadline % The default is
\footline=\paperfootline % \papers
\countdef\pageno=1      \countdef\pagen@=0
\countdef\pagenumber=1  \pagenumber=1
\def\advancepageno{\gl@bal\advance\pagen@ by 1
   \ifnum\pagenumber<0 \gl@bal\advance\pagenumber by -1
    \else\gl@bal\advance\pagenumber by 1 \fi
    \gl@bal\frontpagefalse  \swing@ }
\def\folio{\ifnum\pagenumber<0 \romannumeral-\pagenumber
           \else \number\pagenumber \fi }
\def\swing@{\ifodd\pagenumber \gl@bal\advance\hoffset by -\HSWING
             \else \gl@bal\advance\hoffset by \HSWING \fi }
\def\footrule{\dimen@=\prevdepth\nointerlineskip
   \vbox to 0pt{\vskip -0.25\baselineskip \hrule width 0.35\hsize \vss}
   \prevdepth=\dimen@ }
\let\footnotespecial=\rel@x
\newdimen\footindent
\footindent=24pt
\def\Textindent#1{\noindent\llap{#1\enspace}\ignorespaces}
\def\Vfootnote#1{\insert\footins\bgroup
   \interlinepenalty=\interfootnotelinepenalty \floatingpenalty=20000
   \singl@true\doubl@false\Tenpoint
   \splittopskip=\ht\strutbox \boxmaxdepth=\dp\strutbox
   \leftskip=\footindent \rightskip=\z@skip
   \parindent=0.5\footindent \parfillskip=0pt plus 1fil
   \spaceskip=\z@skip \xspaceskip=\z@skip \footnotespecial
   \Textindent{#1}\footstrut\futurelet\next\fo@t}

\def\vfootnote#1{\Vfootnote{${#1}$}}
\def\footnote#1{\attach{#1}\vfootnote{#1}}

\let\footsymbol=\star
\newcount\lastf@@t           \lastf@@t=-1
\newcount\footsymbolcount    \footsymbolcount=0
\newif\ifPhysRev
\def\bumpfootsymbolcount{\rel@x
   \iffrontpage \bumpfootsymbolpos \else \advance\lastf@@t by 1
     \ifPhysRev \bumpfootsymbolneg \else \bumpfootsymbolpos \fi \fi
   \gl@bal\lastf@@t=\pagen@ }
\def\bumpfootsymbolpos{\ifnum\footsymbolcount <0
                            \gl@bal\footsymbolcount =0 \fi
    \ifnum\lastf@@t<\pagen@ \gl@bal\footsymbolcount=0
     \else \gl@bal\advance\footsymbolcount by 1 \fi }
\def\bumpfootsymbolneg{\ifnum\footsymbolcount >0
             \gl@bal\footsymbolcount =0 \fi
         \gl@bal\advance\footsymbolcount by -1 }
\def\fd@f#1 {\xdef\footsymbol{\mathchar"#1 }}
\def\generatefootsymbol{\ifcase\footsymbolcount \fd@f 13F \or \fd@f 279
        \or \fd@f 27A \or \fd@f 278 \or \fd@f 27B \else
        \ifnum\footsymbolcount <0 \fd@f{023 \number-\footsymbolcount }
         \else \fd@f 203 {\loop \ifnum\footsymbolcount >5
                \fd@f{203 \footsymbol } \advance\footsymbolcount by -1
                \repeat }\fi \fi }

\def\nonfrenchspacing{\sfcode`\.=3001 \sfcode`\!=3000 \sfcode`\?=3000
        \sfcode`\:=2000 \sfcode`\;=1500 \sfcode`\,=1251 }
\nonfrenchspacing
\newdimen\d@twidth
{\setbox0=\hbox{s.} \gl@bal\d@twidth=\wd0 \setbox0=\hbox{s}
        \gl@bal\advance\d@twidth by -\wd0 }
\def\removehglue{\loop \unskip \ifdim\lastskip >\z@ \repeat }
\def\roll@ver#1{\removehglue \nobreak \count255 =\spacefactor \dimen@=\z@
        \ifnum\count255 =3001 \dimen@=\d@twidth \fi
        \ifnum\count255 =1251 \dimen@=\d@twidth \fi
    \iftwelv@ \kern-\dimen@ \else \kern-0.83\dimen@ \fi
   #1\spacefactor=\count255 }
\def\step@ver#1{\rel@x \ifmmode #1\else \ifhmode
        \roll@ver{${}#1$}\else {\setbox0=\hbox{${}#1$}}\fi\fi }
\def\attach#1{\step@ver{\strut^{\mkern 2mu #1} }}
%
%%%%%%%%%%%%%%%%%%%%%%%%%%%%%%%%%%%%%%%%%%%%%%%%%%%%%%%%%%%%%%%%%%%%%%%%
%
%   Here come chapter, section, subsection & appendix macros.
%
\newcount\chapternumber      \chapternumber=0
\newcount\sectionnumber      \sectionnumber=0
\newcount\equanumber         \equanumber=0
\let\chapterlabel=\rel@x
\let\sectionlabel=\rel@x
\newtoks\chapterstyle        \chapterstyle={\Number}
\newtoks\sectionstyle        \sectionstyle={\Number}
\newskip\chapterskip         \chapterskip=\bigskipamount
\newskip\sectionskip         \sectionskip=\medskipamount
\newskip\headskip            \headskip=8pt plus 3pt minus 3pt
\newdimen\chapterminspace    \chapterminspace=15pc
\newdimen\sectionminspace    \sectionminspace=10pc
\newdimen\referenceminspace  \referenceminspace=20pc
\newif\ifcn@                 \cn@true
\newif\ifcn@@                \cn@@false
\def\numberedchapters{\cn@true}
\def\unnumberedchapters{\cn@false\sequentialequations}
\def\chapterreset{\gl@bal\advance\chapternumber by 1
   \ifnum\equanumber<0 \else\gl@bal\equanumber=0\fi
   \sectionnumber=0 \let\sectionlabel=\rel@x
   \ifcn@ \gl@bal\cn@@true {\pr@tect
       \xdef\chapterlabel{\the\chapterstyle{\the\chapternumber}}}%
    \else \gl@bal\cn@@false \gdef\chapterlabel{\rel@x}\fi }
\def\@alpha#1{\count255='140 \advance\count255 by #1\char\count255}
 \def\alphabetic{\n@expand\@alpha}
\def\@Alpha#1{\count255='100 \advance\count255 by #1\char\count255}
 \def\Alphabetic{\n@expand\@Alpha}
\def\@Roman#1{\uppercase\expandafter{\romannumeral #1}}
 \def\Roman{\n@expand\@Roman}
\def\@roman#1{\romannumeral #1}    \def\roman{\n@expand\@roman}
\def\@number#1{\number #1}         \def\Number{\n@expand\@number}
\def\BLANK#1{\rel@x}               
\def\titleparagraphs{\interlinepenalty=9999
     \leftskip=0.03\hsize plus 0.22\hsize minus 0.03\hsize
     \rightskip=\leftskip \parfillskip=0pt
     \hyphenpenalty=9000 \exhyphenpenalty=9000
     \tolerance=9999 \pretolerance=9000
     \spaceskip=0.333em \xspaceskip=0.5em }
\def\titlestyle#1{\par\begingroup \titleparagraphs
     \iftwelv@\fourteenpoint\else\twelvepoint\fi
   \noindent #1\par\endgroup }
\def\spacecheck#1{\dimen@=\pagegoal\advance\dimen@ by -\pagetotal
   \ifdim\dimen@<#1 \ifdim\dimen@>0pt \vfil\break \fi\fi}
\def\chapter#1{\par \penalty-300 \vskip\chapterskip
   \spacecheck\chapterminspace
   \chapterreset \titlestyle{\ifcn@@\chapterlabel.~\fi #1}
   \nobreak\vskip\headskip \penalty 30000
   {\pr@tect\wlog{\string\chapter\space \chapterlabel}} }

\def\section#1{\par \ifnum\lastpenalty=30000\else
   \penalty-200\vskip\sectionskip \spacecheck\sectionminspace\fi
   \gl@bal\advance\sectionnumber by 1
 {\pr@tect
   \xdef\sectionlabel{\ifcn@@ \chapterlabel.\fi
       \the\sectionstyle{\the\sectionnumber}%
                     }%
   \wlog{\string\section\space \sectionlabel}
 }%
   \noindent {\caps\enspace\sectionlabel.~~#1}\par
   \nobreak\vskip\headskip \penalty 30000 }
\def\subsection#1{\par
   \ifnum\the\lastpenalty=30000\else \penalty-100\smallskip \fi
   \noindent\undertext{#1}\enspace \vadjust{\penalty5000}}

\def\undertext#1{\vtop{\hbox{#1}\kern 1pt \hrule}}
\def\ACK{\par\penalty-100\medskip \spacecheck\sectionminspace
   \line{\fourteenrm\hfil ACKNOWLEDGEMENTS\hfil}\nobreak\vskip\headskip }
\def\ack{\subsection{Acknowledgements:}}
\def\APPENDIX#1#2{\par\penalty-300\vskip\chapterskip
   \spacecheck\chapterminspace \chapterreset \xdef\chapterlabel{#1}
   \titlestyle{APPENDIX #2} \nobreak\vskip\headskip \penalty 30000
   \wlog{\string\Appendix~\chapterlabel} }
\def\Appendix#1{\APPENDIX{#1}{#1}}
\def\appendix{\APPENDIX{A}{}}
%
%%%%%%%%%%%%%%%%%%%%%%%%%%%%%%%%%%%%%%%%%%%%%%%%%%%%%%%%%%%%%%%%%%%%%%%%
%
%   Here come macros for equation numbering.
%
\def\eqname#1{\rel@x {\pr@tect
  \ifnum\equanumber<0 \xdef#1{{\rm(\number-\equanumber)}}%
     \gl@bal\advance\equanumber by -1
  \else \gl@bal\advance\equanumber by 1
   \xdef#1{{\rm(\ifcn@@ \chapterlabel.\fi \number\equanumber)}}\fi
  }#1}

\def\eqn{\eqno\eqname}

\def\eqinsert#1{\noalign{\dimen@=\prevdepth \nointerlineskip
   \setbox0=\hbox to\displaywidth{\hfil #1}
   \vbox to 0pt{\kern 0.5\baselineskip\hbox{$\!\box0\!$}\vss}
   \prevdepth=\dimen@}}
%

%
%%%%%%%%%%%%%%%%%%%%%%%%%%%%%%%%%%%%%%%%%%%%%%%%%%%%%%%%%%%%%%%%%%%%%%%%
%   Here come items and lists
%
\def\GENITEM#1;#2{\par \hangafter=0 \hangindent=#1
    \Textindent{$ #2 $}\ignorespaces}
\outer\def\newitem#1=#2;{\gdef#1{\GENITEM #2;}}

\newdimen\itemsize                \itemsize=30pt
\newitem\item=1\itemsize;
\newitem\sitem=1.75\itemsize;     
\newitem\ssitem=2.5\itemsize;     
\outer\def\newlist#1=#2&#3&#4;{\toks0={#2}\toks1={#3}%
   \count255=\escapechar \escapechar=-1
   \alloc@0\list\countdef\insc@unt\listcount     \listcount=0
   \edef#1{\par
      \countdef\listcount=\the\allocationnumber
      \advance\listcount by 1
      \hangafter=0 \hangindent=#4
      \Textindent{\the\toks0{\listcount}\the\toks1}}
   \expandafter\expandafter\expandafter
    \edef\c@t#1{begin}{\par
      \countdef\listcount=\the\allocationnumber \listcount=1
      \hangafter=0 \hangindent=#4
      \Textindent{\the\toks0{\listcount}\the\toks1}}
   \expandafter\expandafter\expandafter
    \edef\c@t#1{con}{\par \hangafter=0 \hangindent=#4 \noindent}
   \escapechar=\count255}
\def\c@t#1#2{\csname\string#1#2\endcsname}
\newlist\point=\Number&.&1.0\itemsize;
\newlist\subpoint=(\alphabetic&)&1.75\itemsize;
\newlist\subsubpoint=(\roman&)&2.5\itemsize;
%

%
%%%%%%%%%%%%%%%%%%%%%%%%%%%%%%%%%%%%%%%%%%%%%%%%%%%%%%%%%%%%%%%%%%%%%%%%
%
%   Here come macros for references, figures & tables.
%
% % % % % % % % % % % % % % % % % % % % % % % % % % % % % % % % % % % %
%%  First, references.
%
\newcount\referencecount     \referencecount=0
\newcount\lastrefsbegincount \lastrefsbegincount=0
\newif\ifreferenceopen       \newwrite\referencewrite
\newdimen\refindent          \refindent=30pt
\def\normalrefmark#1{\attach{\scriptscriptstyle [ #1 ] }}
\let\PRrefmark=\attach
\def\NPrefmark#1{\step@ver{{\;[#1]}}}
\def\refmark#1{\rel@x\ifPhysRev\PRrefmark{#1}\else\normalrefmark{#1}\fi}
\def\refend@{\refmark{\number\referencecount}}
\def\refend{\refend@{}\space }
\def\refsend{\refmark{\count255=\referencecount
   \advance\count255 by-\lastrefsbegincount
   \ifcase\count255 \number\referencecount
   \or \number\lastrefsbegincount,\number\referencecount
   \else \number\lastrefsbegincount-\number\referencecount \fi}\space }
\def\REFNUM#1{\rel@x \gl@bal\advance\referencecount by 1
    \xdef#1{\the\referencecount }}
\def\Refnum#1{\REFNUM #1\refend@ } 
\def\REF#1{\REFNUM #1\R@FWRITE\ignorespaces}
\def\Ref#1{\Refnum #1\REFWRITE }
\def\ref{\Ref\?}
\def\REFS#1{\REFNUM #1\gl@bal\lastrefsbegincount=\referencecount
    \REFWRITE }
\def\refs{\REFS\?}
\def\refc{\REF\?}
\let\refscon=\refc       \let\REFSCON=\REF
\def\r@fitem#1{\par \hangafter=0 \hangindent=\refindent \Textindent{#1}}
\def\refitem#1{\r@fitem{#1.}}
\def\NPrefitem#1{\r@fitem{[#1]}}
\def\NPrefs{\let\refmark=\NPrefmark \let\refitem=\NPrefitem}
\def\REFWRITE{\R@FWRITE\rel@x }
\def\R@FWRITE#1{\ifreferenceopen \else \gl@bal\referenceopentrue
     \immediate\openout\referencewrite=\jobname.refs
     \toks@={\begingroup \refoutspecials \catcode`\^^M=10 }%
     \immediate\write\referencewrite{\the\toks@}\fi
    \immediate\write\referencewrite{\noexpand\refitem %
                                    {\the\referencecount}}%
    \p@rse@ndwrite \referencewrite #1}
\begingroup
 \catcode`\^^M=\active \let^^M=\relax %
 \gdef\p@rse@ndwrite#1#2{\begingroup \catcode`\^^M=12 \newlinechar=`\^^M%
         \chardef\rw@write=#1\sc@nlines#2}%
 \gdef\sc@nlines#1#2{\sc@n@line \g@rbage #2^^M\endsc@n \endgroup #1}%
 \gdef\sc@n@line#1^^M{\expandafter\toks@\expandafter{\deg@rbage #1}%
         \immediate\write\rw@write{\the\toks@}%
         \futurelet\n@xt \sc@ntest }%
\endgroup
\def\sc@ntest{\ifx\n@xt\endsc@n \let\n@xt=\rel@x
       \else \let\n@xt=\sc@n@notherline \fi \n@xt }
\def\sc@n@notherline{\sc@n@line \g@rbage }
\def\deg@rbage#1{}
\let\g@rbage=\relax    \let\endsc@n=\relax
\def\refout{\par\penalty-400\vskip\chapterskip
   \spacecheck\referenceminspace
   \ifreferenceopen \Closeout\referencewrite \referenceopenfalse \fi
   \line{\fourteenrm\hfil REFERENCES\hfil}\vskip\headskip
   \input \jobname.refs
   }
\def\refoutspecials{\sfcode`\.=1000 \interlinepenalty=1000
         \rightskip=\z@ plus 1em minus \z@ }
\def\Closeout#1{\toks0={\par\endgroup}\immediate\write#1{\the\toks0}%
   \immediate\closeout#1}
%
% % % % % % % % % % % % % % % % % % % % % % % % % % % % % % % % % % % %
%%  Next, figure captions and table captions.
%
\newcount\figurecount     \figurecount=0
\newcount\tablecount      \tablecount=0
\newif\iffigureopen       \newwrite\figurewrite
\newif\iftableopen        \newwrite\tablewrite
\def\FIGNUM#1{\rel@x \gl@bal\advance\figurecount by 1
    \xdef#1{\the\figurecount}}
\def\FIGURE#1{\FIGNUM #1\F@GWRITE\ignorespaces }

\def\figitem#1{\r@fitem{#1)}}
\def\FIGWRITE{\F@GWRITE\rel@x }
\def\TABNUM#1{\rel@x \gl@bal\advance\tablecount by 1
    \xdef#1{\the\tablecount}}
\def\TABLE#1{\TABNUM #1\T@BWRITE\ignorespaces }

\def\tabitem#1{\r@fitem{#1:}}
\def\TABWRITE{\T@BWRITE\rel@x }
\def\F@GWRITE#1{\iffigureopen \else \gl@bal\figureopentrue
     \immediate\openout\figurewrite=\jobname.figs
     \toks@={\begingroup \catcode`\^^M=10 }%
     \immediate\write\figurewrite{\the\toks@}\fi
    \immediate\write\figurewrite{\noexpand\figitem %
                                 {\the\figurecount}}%
    \p@rse@ndwrite \figurewrite #1}
\def\T@BWRITE#1{\iftableopen \else \gl@bal\tableopentrue
     \immediate\openout\tablewrite=\jobname.tabs
     \toks@={\begingroup \catcode`\^^M=10 }%
     \immediate\write\tablewrite{\the\toks@}\fi
    \immediate\write\tablewrite{\noexpand\tabitem %
                                 {\the\tablecount}}%
    \p@rse@ndwrite \tablewrite #1}
\def\figout{\par\penalty-400
   \vskip\chapterskip\spacecheck\referenceminspace
   \iffigureopen \Closeout\figurewrite \figureopenfalse \fi
   \line{\fourteenrm\hfil FIGURE CAPTIONS\hfil}\vskip\headskip
   \input \jobname.figs
   }
\def\tabout{\par\penalty-400
   \vskip\chapterskip\spacecheck\referenceminspace
   \iftableopen \Closeout\tablewrite \tableopenfalse \fi
   \line{\fourteenrm\hfil TABLE CAPTIONS\hfil}\vskip\headskip
   \input \jobname.tabs
   }
%
% % % % % % % % % % % % % % % % % % % % % % % % % % % % % % % % % % % %
%%  Finally, inserted figures.
%
%
\newbox\picturebox
\def\p@cht{\ht\picturebox }
\def\p@cwd{\wd\picturebox }
\def\p@cdp{\dp\picturebox }
\newdimen\xshift
\newdimen\yshift
\newdimen\captionwidth
\newskip\captionskip
\captionskip=15pt plus 5pt minus 3pt
\def\fullwidth{\captionwidth=\hsize }
\newtoks\Caption
\newif\ifcaptioned
\newif\ifselfcaptioned
\def\caption{\captionedtrue \Caption }
\newcount\linesabove
\newif\iffileexists
\newtoks\picfilename
\def\fil@#1 {\fileexiststrue \picfilename={#1}}
\def\file#1{\if=#1\let\n@xt=\fil@ \else \def\n@xt{\fil@ #1}\fi \n@xt }
\def\pl@t{\begingroup \pr@tect
    \setbox\picturebox=\hbox{}\fileexistsfalse
    \let\height=\p@cht \let\width=\p@cwd \let\depth=\p@cdp
    \xshift=\z@ \yshift=\z@ \captionwidth=\z@
    \Caption={}\captionedfalse
    \linesabove =0 \picturedefault }
\def\plot{\pl@t \selfcaptionedfalse }
\def\Picture#1{\gl@bal\advance\figurecount by 1
    \xdef#1{\the\figurecount}\pl@t \selfcaptionedtrue }

\def\s@vepicture{\iffileexists \parsefilename \redopicturebox \fi
   \ifdim\captionwidth>\z@ \else \captionwidth=\p@cwd \fi
   \xdef\lastpicture{\iffileexists
        \setbox0=\hbox{\raise\the\yshift \vbox{%
              \moveright\the\xshift\hbox{\picturedefinition}}}%
        \else \setbox0=\hbox{}\fi
         \ht0=\the\p@cht \wd0=\the\p@cwd \dp0=\the\p@cdp
         \vbox{\hsize=\the\captionwidth \line{\hss\box0 \hss }%
              \ifcaptioned \vskip\the\captionskip \noexpand\Tenpoint
                \ifselfcaptioned Figure~\the\figurecount.\enspace \fi
                \the\Caption \fi }}%
    \endgroup }
\let\endpicture=\s@vepicture
\def\savepicture#1{\s@vepicture \global\let#1=\lastpicture }
\def\displaypicture{\fullwidth \s@vepicture $$\lastpicture $${}}
\def\toppicture{\fullwidth \s@vepicture \topinsert
    \lastpicture \medskip \endinsert }
\def\midpicture{\fullwidth \s@vepicture \midinsert
    \lastpicture \endinsert }
%
%  Wraparound macros - a try.
%
\def\leftpicture{\pres@tpicture
    \dimen@i=\hsize \advance\dimen@i by -\dimen@ii
    \setbox\picturebox=\hbox to \hsize {\box0 \hss }%
    \wr@paround }
\def\rightpicture{\pres@tpicture
    \dimen@i=\z@
    \setbox\picturebox=\hbox to \hsize {\hss \box0 }%
    \wr@paround }
\def\pres@tpicture{\gl@bal\linesabove=\linesabove
    \s@vepicture \setbox\picturebox=\vbox{
         \kern \linesabove\baselineskip \kern 0.3\baselineskip
         \lastpicture \kern 0.3\baselineskip }%
    \dimen@=\p@cht \dimen@i=\dimen@
    \advance\dimen@i by \pagetotal
    \par \ifdim\dimen@i>\pagegoal \vfil\break \fi
    \dimen@ii=\hsize
    \advance\dimen@ii by -\parindent \advance\dimen@ii by -\p@cwd
    \setbox0=\vbox to\z@{\kern-\baselineskip \unvbox\picturebox \vss }}
\def\wr@paround{\Caption={}\count255=1
    \loop \ifnum \linesabove >0
         \advance\linesabove by -1 \advance\count255 by 1
         \advance\dimen@ by -\baselineskip
         \expandafter\Caption \expandafter{\the\Caption \z@ \hsize }%
      \repeat
    \loop \ifdim \dimen@ >\z@
         \advance\count255 by 1 \advance\dimen@ by -\baselineskip
         \expandafter\Caption \expandafter{%
             \the\Caption \dimen@i \dimen@ii }%
      \repeat
    \edef\n@xt{\parshape=\the\count255 \the\Caption \z@ \hsize }%
    \par\noindent \n@xt \strut \vadjust{\box\picturebox }}
\let\picturedefault=\relax
\let\parsefilename=\relax
\def\redopicturebox{\let\picturedefinition=\rel@x
   \errhelp=\disabledpictures
   \errmessage{This version of TeX cannot handle pictures.  Sorry.}}
\newhelp\disabledpictures
     {You will get a blank box in place of your picture.}
%
%
%
% Above definitions of \parsefilename and \redopicturebox
% are dumb defaults.  Actual definition are system dependent;
% you will probably find them in your PHYZZX.LOCAL file.
%
% The example below is used at Princeton.
%
%\def\parsefilename{\expandafter\p@rse\the\picfilename.\endp@rse }
%\def\p@rse#1.#2\endp@rse{\if"#2"\expandafter\picfilename
%        \expandafter{\the\picfilename.fig}\fi }
%
%\newread\pictureread
%\def\redopicturebox{\expandafter\openin\expandafter\pictureread
%                    \the\picfilename
%   \ifeof\pictureread \errhelp=\disabledpictures
%     \edef\n@xt{\errmessage={Cannot find file \the\picfilename}\n@xt
%     \let\pictureboxdefinition=\relax \else
%    \read\pictureread to\y@p \read\pictureread to\y@p
%    \read\pictureread to\x@p \read\pictureread to\y@m
%    \read\pictureread to\x@m \closein\pictureread
%    \p@cht=\y@p truein \advance\p@cht by -\y@m truein
%    \advance\yshift by \y@p truein
%    \p@cwd=\x@p truein \advance\p@cwd by -\x@m truein
%    \advance\xshift by \x@m truein
%    \edef\picturedefinition{\special{pos,inc=\the\picfilename}}%
%    \fi }
%
%
%%%%%%%%%%%%%%%%%%%%%%%%%%%%%%%%%%%%%%%%%%%%%%%%%%%%%%%%%%%%%%%%%%%%%%%%
%
%   Here come macros for memos & letters.
%
\def\FRONTPAGE{\ifvoid255\else\vfill\penalty-20000\fi
   \gl@bal\pagenumber=1     \gl@bal\chapternumber=0
   \gl@bal\equanumber=0     \gl@bal\sectionnumber=0
   \gl@bal\referencecount=0 \gl@bal\figurecount=0
   \gl@bal\tablecount=0     \gl@bal\frontpagetrue
   \gl@bal\lastf@@t=0       \gl@bal\footsymbolcount=0
   \gl@bal\cn@@false }

\def\papers{\papersize\headline=\paperheadline\footline=\paperfootline}
\def\papersize{\hsize=35pc \vsize=50pc \hoffset=0pc \voffset=1pc
   \advance\hoffset by\HOFFSET \advance\voffset by\VOFFSET
   \pagebottomfiller=0pc
   \skip\footins=\bigskipamount \normalspace }
\papers  %  This is the default
%
% % % % % % % % % % % % % % % % % % % % % % % % % % % % % % % % % % % %
%
\newskip\lettertopskip       \lettertopskip=20pt plus 50pt
\newskip\letterbottomskip    \letterbottomskip=\z@ plus 100pt
\newskip\signatureskip       \signatureskip=40pt plus 3pt
\def\lettersize{\hsize=6.5in \vsize=8.5in \hoffset=0in \voffset=0.5in
   \advance\hoffset by\HOFFSET \advance\voffset by\VOFFSET
   \pagebottomfiller=\letterbottomskip
   \skip\footins=\smallskipamount \multiply\skip\footins by 3
   \singlespace }
\def\MEMO{\lettersize \headline=\letterheadline \footline={\hfil }%
   \let\rule=\memorule \FRONTPAGE \memohead }

\def\memodate{\afterassignment\MEMO \date }
\def\memit@m#1{\smallskip \hangafter=0 \hangindent=1in
    \Textindent{\caps #1}}
\def\subject{\memit@m{Subject:}}
\def\topic{\memit@m{Topic:}}
\def\from{\memit@m{From:}}
\def\to{\rel@x \ifmmode \rightarrow \else \memit@m{To:}\fi }
\def\memorule{\medskip\hrule height 1pt\bigskip}  % default definitions
\def\memohead{\centerline{\fourteenrm MEMORANDUM}}% see phyzzx.local
\newwrite\labelswrite
\newtoks\rw@toks
\def\letters{\lettersize
   \headline=\letterheadline \footline=\letterfootline
   \immediate\openout\labelswrite=\jobname.lab}

\let\letterhead=\rel@x
\def\addressee#1{\medskip\line{\hskip 0.75\hsize plus\z@ minus 0.25\hsize
                               \the\date \hfil }%
   \vskip \lettertopskip
   \ialign to\hsize{\strut ##\hfil\tabskip 0pt plus \hsize \crcr #1\crcr}
   \writelabel{#1}\medskip \noindent\hskip -\spaceskip \ignorespaces }
\def\rwl@begin#1\cr{\rw@toks={#1\crcr}\rel@x
   \immediate\write\labelswrite{\the\rw@toks}\futurelet\n@xt\rwl@next}
\def\rwl@next{\ifx\n@xt\rwl@end \let\n@xt=\rel@x
      \else \let\n@xt=\rwl@begin \fi \n@xt}
\let\rwl@end=\rel@x
\def\writelabel#1{\immediate\write\labelswrite{\noexpand\labelbegin}
     \rwl@begin #1\cr\rwl@end
     \immediate\write\labelswrite{\noexpand\labelend}}
\newtoks\FromAddress         \FromAddress={}
\newtoks\sendername          \sendername={}
\newbox\FromLabelBox
\newdimen\labelwidth          \labelwidth=6in
\def\makelabels{\afterassignment\Makelabels \sendername=}
\def\Makelabels{\FRONTPAGE \letterinfo={\hfil } \MakeFromBox
     \immediate\closeout\labelswrite  \input \jobname.lab\vfil\eject}
\let\labelend=\rel@x
\def\labelbegin#1\labelend{\setbox0=\vbox{\ialign{##\hfil\cr #1\crcr}}
     \MakeALabel }
\def\MakeFromBox{\gl@bal\setbox\FromLabelBox=\vbox{\Tenpoint
     \ialign{##\hfil\cr \the\sendername \the\FromAddress \crcr }}}
\def\MakeALabel{\vskip 1pt \hbox{\vrule \vbox{
        \hsize=\labelwidth \hrule\bigskip
        \leftline{\hskip 1\parindent \copy\FromLabelBox}\bigskip
        \centerline{\hfil \box0 } \bigskip \hrule
        }\vrule } \vskip 1pt plus 1fil }
\def\signed#1{\par \nobreak \bigskip \dt@pfalse \begingroup
  \everycr={\noalign{\nobreak
            \ifdt@p\vskip\signatureskip\gl@bal\dt@pfalse\fi }}%
  \tabskip=0.5\hsize plus \z@ minus 0.5\hsize
  \halign to\hsize {\strut ##\hfil\tabskip=\z@ plus 1fil minus \z@\crcr
          \noalign{\gl@bal\dt@ptrue}#1\crcr }%
  \endgroup \bigskip }
\newbox\letterb@x
\def\lettertext{\par \vskip\parskip \unvcopy\letterb@x \par }
\def\multiletter{\setbox\letterb@x=\vbox\bgroup
      \everypar{\vrule height 1\baselineskip depth 0pt width 0pt }
      \singlespace \topskip=\baselineskip }
\def\letterend{\par\egroup}
%
%%%%%%%%%%%%%%%%%%%%%%%%%%%%%%%%%%%%%%%%%%%%%%%%%%%%%%%%%%%%%%%%%%%%%%%
%
%   Here come macros for title pages.
%
\newskip\frontpageskip
\newtoks\Pubnum   
\newtoks\Pubtype  \let\pubtype=\Pubtype
\newif\ifp@bblock  \p@bblocktrue
\def\PH@SR@V{\doubl@true \baselineskip=24.1pt plus 0.2pt minus 0.1pt
             \parskip= 3pt plus 2pt minus 1pt }
\def\PHYSREV{\papers\PhysRevtrue\PH@SR@V}

\def\titlepage{\FRONTPAGE\papers\ifPhysRev\PH@SR@V\fi
   \ifp@bblock\p@bblock \else\hrule height\z@ \rel@x \fi }
\def\nopubblock{\p@bblockfalse}

\frontpageskip=12pt plus .5fil minus 2pt
\Pubtype={}
\Pubnum={}
\def\p@bblock{\begingroup \tabskip=\hsize minus \hsize
   \baselineskip=1.5\ht\strutbox \topspace-2\baselineskip
   \halign to\hsize{\strut ##\hfil\tabskip=0pt\crcr
       \the\Pubnum\crcr\the\date\crcr\the\pubtype\crcr}\endgroup}
\def\title#1{\vskip\frontpageskip \titlestyle{#1} \vskip\headskip }
\def\author#1{\vskip\frontpageskip\titlestyle{\twelvecp #1}\nobreak}

\def\address#1{\par\kern 5pt\titlestyle{\twelvepoint\it #1}}
\def\andaddress{\par\kern 5pt \centerline{\sl and} \address}

\def\abstract{\par\dimen@=\prevdepth \hrule height\z@ \prevdepth=\dimen@
   \vskip\frontpageskip\centerline{\fourteenrm ABSTRACT}\vskip\headskip }

%
%
%%%%%%%%%%%%%%%%%%%%%%%%%%%%%%%%%%%%%%%%%%%%%%%%%%%%%%%%%%%%%%%%%%%%%%%%
%   Miscellaneous macros
%

\def\\{\rel@x \ifmmode \backslash \else {\tt\char`\\}\fi }
\def\sequentialequations{\rel@x \if\equanumber<0 \else
  \gl@bal\equanumber=-\equanumber \gl@bal\advance\equanumber by -1 \fi }
\def\journal#1&#2(#3){\begingroup \let\journal=\dummyj@urnal
    \unskip, \sl #1\unskip~\bf\ignorespaces #2\rm
    (\afterassignment\j@ur \count255=#3), \endgroup\ignorespaces }
\def\j@ur{\ifnum\count255<100 \advance\count255 by 1900 \fi
          \number\count255 }
\def\dummyj@urnal{%
    \toks@={Reference foul up: nested \journal macros}%
    \errhelp={Your forgot & or ( ) after the last \journal}%
    \errmessage{\the\toks@ }}

\def\topspace{\hrule height 0pt depth 0pt \vskip}

\def\Buildrel#1\under#2{\mathrel{\mathop{#2}\limits_{#1}}}
\def\becomes#1{\mathchoice{\becomes@\scriptstyle{#1}}
   {\becomes@\scriptstyle{#1}} {\becomes@\scriptscriptstyle{#1}}
   {\becomes@\scriptscriptstyle{#1}}}
\def\becomes@#1#2{\mathrel{\setbox0=\hbox{$\m@th #1{\,#2\,}$}%
        \mathop{\hbox to \wd0 {\rightarrowfill}}\limits_{#2}}}

\let\int=\intop         
\def\lsim{\mathrel{\mathpalette\@versim<}}
\def\gsim{\mathrel{\mathpalette\@versim>}}
\def\@versim#1#2{\vcenter{\offinterlineskip
        \ialign{$\m@th#1\hfil##\hfil$\crcr#2\crcr\sim\crcr } }}
\def\big#1{{\hbox{$\left#1\vbox to 0.85\b@gheight{}\right.\n@space$}}}
\def\Big#1{{\hbox{$\left#1\vbox to 1.15\b@gheight{}\right.\n@space$}}}
\def\bigg#1{{\hbox{$\left#1\vbox to 1.45\b@gheight{}\right.\n@space$}}}
\def\Bigg#1{{\hbox{$\left#1\vbox to 1.75\b@gheight{}\right.\n@space$}}}
\def\){\mskip 2mu\nobreak }
%
% % % % % % % % % % % % % % % % % % % % % % % % % % % % % % % % % % % %
%
%   Finally, some bug fixings.
%
\let\sec@nt=\sec
\def\sec{\rel@x\ifmmode\let\n@xt=\sec@nt\else\let\n@xt\section\fi\n@xt}
\def\obsolete#1{\message{Macro \string #1 is obsolete.}}
\def\firstsec#1{\obsolete\firstsec \section{#1}}
\def\firstsubsec#1{\obsolete\firstsubsec \subsection{#1}}
\def\thispage#1{\obsolete\thispage \gl@bal\pagenumber=#1\frontpagefalse}
\def\thischapter#1{\obsolete\thischapter \gl@bal\chapternumber=#1}
\def\splitout{\obsolete\splitout\rel@x}
\def\prop{\obsolete\prop \propto }
\def\nextequation#1{\obsolete\nextequation \gl@bal\equanumber=#1
   \ifnum\the\equanumber>0 \gl@bal\advance\equanumber by 1 \fi}
\def\BOXITEM{\afterassigment\B@XITEM\setbox0=}
\def\B@XITEM{\par\hangindent\wd0 \noindent\box0 }
%
%
%%%%%%%%%%%%%%%%%%%%%%%%%%%%%%%%%%%%%%%%%%%%%%%%%%%%%%%%%%%%%%%%%%%%%%%%
%   That's about it
%
\def\phyzzx{PHY\setbox0=\hbox{Z}\copy0 \kern-0.5\wd0 \box0 X}
        
\everyjob{\xdef\today{\monthname~\number\day, \number\year}
        \input myphyx.tex }
\message{ by V.K.}
%
%%%%%%%%%%%%%%%%%%%%%%%%%%%%%%%%%%%%%%%%%%%%%%%%%%%%%%%%%%%%%%%%%%%%%%%%
%%%                 This is PHYZZX.LOCAL (cm version)                %%%
%%%%%%%%%%%%%%%%%%%%%%%%%%%%%%%%%%%%%%%%%%%%%%%%%%%%%%%%%%%%%%%%%%%%%%%%
%%%%%%%     Latest update/debug: June 24, 1992.         %%%%%%%%%%%%%%%%
%%%%%%%%%%%%%%%%%%%%%%%%%%%%%%%%%%%%%%%%%%%%%%%%%%%%%%%%%%%%%%%%%%%%%%%%
%
%
%
%
%	Modifications to PHYZZX to make it UW specific
% 	-includes: a new memo header
% 		   a new letter head
% 		   a UW HEP publication number
%                  a new label output routine for use with MAKELABELS.TEX
%
%    ADD SOME MORE FONTS
%\font\twentyrm=cmr10  at 20truept %scaled\magstep4

%
%
%\def\MEMO{\letterstyle\FRONTPAGE \letterfrontheadline={\hfil}
%    \line{\quad\fourteenss MEMORANDUM\hfil\twelvess\the\date\quad}
%    \medskip \memod@f}
%\def\MEMOT{\letterstyle\FRONTPAGE \MADAFHEAD\letterfrontheadline={\hfil}
%    \line{\quad\fourteenss MEMORANDUM\hfil\twelvess\the\date\quad}
%    \medskip \memod@f}
%\def\MEMOX{\letterstyle\FRONTPAGE \MADPHHEAD\letterfrontheadline={\hfil}
%    \line{\quad\fourteenss MEMORANDUM\hfil\twelvess\the\date\quad}
%    \medskip \memod@f}
\showboxbreadth=1000 %
\showboxdepth=5
\def\figitem#1{\r@fitem{#1.}}
\def\tabitem#1{\r@fitem{#1.}}

\def\sequentialequations{\rel@x \ifnum\equanumber<0 \else
  \gl@bal\equanumber=-\equanumber \gl@bal\advance\equanumber by -1 \fi }

\def\boxit#1{\vbox{\hrule\hbox{\vrule\kern3pt
\vbox{\kern3pt#1\kern3pt}\kern3pt\vrule}\hrule}}
%
%
%%%%%%%%%%%%%%%%%%%%%%%%%%%%%%%%%%%%%%%%%%%%%%%%%%%%%%%%%%%%%%%%%%%%%%%%%%%
%%
%%            Making double-column  (these are modified from manmac.tex)
%%            with a full size columns as well.
%%            (This is still buggy--gives overfull boxes etc.)
%%            Report bugs to T J Allen (tjallen@wishep.physics.wisc.edu,
%%            tjallen@suhep.phy.syr.edu or  tja@theory3.caltech.edu)
%%
%%            This will NOT produce double columns in preprintmode since
%%            there are conflicting \output commands.  The whole
%%            macro should be rewritten using a modified \output.
%%
%%            Where you want
%%            the doublecolumn output to start, use \begindoublecolumns.
%%            Where you want to go back to single columns use
%%            \enddoublecolumns.  This produces output much like that
%%            of RevTeX.  If you wish to specify that there be a rule
%%            between the columns of output, then set \columnrulewidth
%%            = 0.4pt.
%%
%%%%%%%%%%%%%%%%%%%%%%%%%%%%%%%%%%%%%%%%%%%%%%%%%%%%%%%%%%%%%%%%%%%%%%
%%
\newbox\partialpage
\newdimen\pageheight \pageheight=\vsize
\newdimen\pagewidth  \pagewidth=6.6truein
\newdimen\columnwidth  \columnwidth=3.2truein
\newdimen\columnrulewidth \columnrulewidth=0pt
\newdimen\ruleht \ruleht=.5pt
\newinsert\margin
\def\twocolumn{%
   \singlespace
   \vsize=9truein
   \pagetextwidth=\pagewidth
   \hsize=\pagewidth
   \titlepagewidth=\pagewidth
   \hoffset=0truein
   \voffset=0truein
   \dimen\margin=\maxdimen
   \count\margin=0 \skip\margin=0pt
 \def\begindoublecolumns{
     \ifpr@printstyle
     \message{ I'm unable to print double columns in PREPRINTSTYLE }
     \end\fi
     \begingroup
     \global\vsize=2\pageheight
     \output={\global\setbox\partialpage=\vbox{\unvbox255\bigskip\bigskip}
         \global\vsize=2\pageheight\global\advance\vsize by -2\ht\partialpage
         \global\advance\vsize by 2\bigskipamount
         \global\advance\vsize by 1 pc}\eject % a little extra room; 1pc
     \output={\doublecolumnout\global\vsize=2\pageheight}
         \global\pagetextwidth=\columnwidth \global\hsize=\columnwidth}
% keeps footnotes on correct page
%
  \def\enddoublecolumns{\output={\balancecolumns\global\hsize=\pagewidth
                       \global\pagetextwidth=\pagewidth
                       \global\vsize=\pageheight
                       \unvbox255 }\eject\endgroup}
  \def\doublecolumnout{\splittopskip=\topskip \splitmaxdepth=\maxdepth
     \dimen@=\pageheight\advance\dimen@ by -\ht\partialpage
     \setbox0=\vsplit255 to\dimen@ \setbox2=\vsplit255 to \dimen@
     \onepageout\pagesofar \unvbox255 \penalty\outputpenalty}
  \def\pagesofar{\unvbox\partialpage
     \wd0=\columnwidth \wd2=\columnwidth \hbox to \pagewidth{\box0\hfil
     \columnrule \hfil \box2}}
  \def\columnrule{\vrule width \columnrulewidth height \ht2}
  \def\balancecolumns{\setbox0=\vbox{\unvbox255}\dimen@=\ht0
     \advance\dimen@ by \topskip \advance\dimen@ by-\baselineskip
     \advance\dimen@ by -2\ht\partialpage  % what if we begin and end on the
     \divide\dimen@ by2                    % same page?!?
     \splittopskip=\topskip
     {\vbadness=10000 \loop \global\setbox3=\copy0
        \global\setbox1=\vsplit3 to \dimen@
        \ifdim\ht3>\dimen@ \global\advance\dimen@ by1pt \repeat}
     \setbox0=\vbox to \dimen@{\unvbox1} \setbox2=\vbox to\dimen@{\dimen2=\dp3
     \unvbox3 \kern-\dimen2 \vfil }
     \pagesofar }
   \def\onepageout##1{ \setbox0=\vbox{##1} \dimen@=\dp0
     \shipout\vbox{ % here we define one page of output
     \makeheadline
     \vbox to \pageheight{
       \boxmaxdepth=\maxdepth
       \ifvoid\margin\else % marginal info is present
       \rlap{\kern31pc\vbox to 0pt{\kern4pt\box\margin\vss}}\fi
       \ifvoid\topins\else\unvbox\topins\vskip\skip\topins\fi
       ##1                                  % now insert the main information
       \vskip\pagebottomfiller
       \ifvoid\footins\else\vskip\skip\footins\footrule\unvbox\footins\fi
       \ifr@ggedbottom\kern-\dimen@ \vfil\fi}  %need a replacement for here
       \makefootline}
     \advancepageno\frontpagefalse}
   \def\makefootline{\baselineskip = 1.5\normalbaselineskip
             \hbox to \pagewidth{\the\footline}}
   \def\footrule{\dimen@=\prevdepth\nointerlineskip
      \vbox to 0pt{\vskip -0.25\baselineskip \hrule width 0.62\pagewidth \vss}
      \prevdepth=\dimen@ }
   \def\Vfootnote##1{\insert\footins\bgroup
      \interlinepenalty=\interfootnotelinepenalty \floatingpenalty=20000
      \singl@true\doubl@false\Tenpoint \hsize=\pagewidth
      \splittopskip=\ht\strutbox \boxmaxdepth=\dp\strutbox
      \leftskip=\footindent \rightskip=\z@skip
      \parindent=0.5\footindent \parfillskip=0pt plus 1fil
      \spaceskip=\z@skip \xspaceskip=\z@skip \footnotespecial
      \Textindent{##1}\footstrut\futurelet\next\fo@t}
% make the footnotes all have the correct size and the same footrule!
%% Make column separators for large one-column equations %%%%%%%%%%%%%%%%%%%
 \def\sp@cecheck##1{\dimen@=\pagegoal\advance\dimen@ by -\pagetotal
      \ifdim\dimen@<##1 \ifdim\dimen@>0pt \vfil\break \fi\fi}
 \def\endleftcolumn{\dimen@=\pagegoal\advance\dimen@ by -\pagetotal
      \ifdim\dimen@<\chapterminspace \ifdim\dimen@>0pt \vfil\break \fi
      \hbox{\vbox{\hrule width \columnwidth}\hbox to 0.4pt
      {\vrule height 10pt depth 0pt}\hfil}\fi}
 \def\beginrightcolumn{\dimen@=\pagegoal\advance\dimen@ by -\pagetotal
      \ifdim\dimen@<\chapterminspace \ifdim\dimen@>0pt \vfil\break \fi
      \hbox to \hsize{\hss\hbox{\vrule height 0pt depth 10pt
      \vbox{\hrule width \columnwidth}}}\fi}
}
%
%
%%%%%%%%%%%%%%%%%%%%%%%%%%%%%%%%%%%%%%%%%%%%%%%%%%%%%%%%%%%%%%%%%%%%%%%%
%
%   Now start the draftmode and preprintmode enhancement features
%   (Homage to harvmac.tex)
%   Report any bugs to T. J. Allen
%       (tjallen@suhep.phy.syr.edu, tja@theory3.caltech.edu and
%        tjallen@wishep.physics.wisc.edu)
%
%%%%%%%%%%%%%%%%%%%%%%%%%%%%%%%%%%%%%%%%%%%%%%%%%%%%%%%%%%%%%%%%%%%%%%%%
%%
%%   Next, I define output routines, footnotes & related stuff.
%%   (The headline has been modified for draftmode and preprints
%%   may be produced in landscape form, two columns sideways)
%%
%%%%%%%%%%%%%%%%%%%%%%%%%%%%%%%%%%%%%%%%%%%%%%%%%%%%%%%%%%%%%%%%%%%%%%
%
%
\newif\ifpr@printstyle \pr@printstylefalse
\newbox\leftpage
\newdimen\fullhsize
\newdimen\titlepagewidth
\newdimen\pagetextwidth
\def\preprintstyle{%
       \message{(This will be printed PREPRINTSTYLE)}
       \let\lr=L
       \frontpagetrue
       \pr@printstyletrue
       \vsize=7truein
       \pagetextwidth=4.75truein
       \fullhsize=10truein
       \titlepagewidth=8truein
       \normalspace
       \Tenpoint
       \voffset=-.31truein
       \hoffset=-.46truein
       \iffrontpage\hsize=\titlepagewidth\else\hsize=\pagetextwidth\fi
 \output={%
    \iffrontpage
      \shipout\vbox{\special{\printertype}\makeheadline
      \hbox to \fullhsize{\hfill\pagebody\hfill}}
      \advancepageno
    \else
       \almostshipout{\leftline{\vbox{\pagebody\makefootline}}}\advancepageno
    \fi}
        \def\almostshipout##1{\if L\lr \count2=1
             \message{[\the\count0.\the\count1.\the\count2]}
        \global\setbox\leftpage=##1 \global\let\lr=R
                             \else \count2=2
        \shipout\vbox{\special{\printertype}
        \hbox to\fullhsize{\hfill\box\leftpage\hskip0.5truein##1\hfill}}
        \global\let\lr=L     \fi}
   \multiply\chapterminspace by 7 \divide\chapterminspace by 9
   \multiply\sectionminspace by 7 \divide\sectionminspace by 9
   \multiply\referenceminspace by 7 \divide\referenceminspace  by 9
   \multiply\chapterskip by 7 \divide\chapterskip  by 9
   \multiply\sectionskip  by 7 \divide\sectionskip  by 9
   \multiply\headskip   by 7 \divide\headskip by 9
   \multiply\baselineskip   by 7 \divide\baselineskip by 9
   \multiply\abovedisplayskip by 7 \divide\abovedisplayskip by 9
   \belowdisplayskip = \abovedisplayskip
\def\advancepageno{\if L\lr \gl@bal\advance\pagen@ by 1\fi
   \ifnum\pagenumber<0 \gl@bal\advance\pagenumber by -1
    \else\gl@bal\advance\pagenumber by 1 \fi
    \gl@bal\frontpagefalse  \swing@
    \gl@bal\hsize=\pagetextwidth}
} % end of preprintstyle specs

\tolerance=1000
\def\printertype{ps: }
%
% Default values for the fullsize document page

\paperheadline={\ifdr@ftmode\hfil\draftdate\else\hfill\fi}
\def\advancepageno{\gl@bal\advance\pagen@ by 1
   \ifnum\pagenumber<0 \gl@bal\advance\pagenumber by -1
    \else\gl@bal\advance\pagenumber by 1 \fi
    \gl@bal\frontpagefalse  \swing@
    \gl@bal\hsize=\pagetextwidth} %MODIFICATION
\def\papersize{\fullhsize=6.5in
               \pagetextwidth=6.5in
               \hsize=\fullhsize
               \vsize=9truein
               \hoffset=0.05 truein
               \voffset=-0.1truein
               \advance\hoffset by\HOFFSET
               \advance\voffset by\VOFFSET
               \pagebottomfiller=0pc
               \skip\footins=\bigskipamount
               \normalspace }
\papers
\def\lettersize{\fullhsize=6.5in
                \pagetextwidth=6.5in
                \hsize=\fullhsize
                \vsize=8.5in
                \hoffset=0in
                \voffset=0.5in
                \advance\hoffset by\HOFFSET
                \advance\voffset by\VOFFSET
                \pagebottomfiller=\letterbottomskip
                \skip\footins=\smallskipamount
                \multiply\skip\footins by 3
                \singlespace }
\def\semi{;\hfil\break}
%%%%%%%%%%%%%%%%%%%%%%%%%%%%%%%%%%%%%%%%%%%%%%%%%%%%%%%%%%%%%%%%%%%%%%%%
%%
%%   Here come chapter, section, subsection & appendix macros.
%%
%%%%%%%%%%%%%%%%%%%%%%%%%%%%%%%%%%%%%%%%%%%%%%%%%%%%%%%%%%%%%%%%%%%%%%%%
%
%  The following allows a shortcut for making titles bold etc.
%  Just use \chapterheadstyle={\bf} in the beginning of the
%  TeX file.
%
\newtoks\chapterheadstyle  \chapterheadstyle={\relax}
\def\chapter#1{{\the\chapterheadstyle\par \penalty-300 \vskip\chapterskip
   \spacecheck\chapterminspace
   \chapterreset \titlestyle{\ifcn@@\chapterlabel.~\fi #1}
   \nobreak\vskip\headskip \penalty 30000
   \message{(\the\chapternumber. #1)}
  {\pr@tect\wlog{\string\chapter\space \chapterlabel}} }}

\def\APPENDIX#1#2{{\the\chapterheadstyle\par\penalty-300\vskip\chapterskip
   \spacecheck\chapterminspace \chapterreset \xdef\chapterlabel{#1}
   \titlestyle{APPENDIX #2} \nobreak\vskip\headskip \penalty 30000
   \wlog{\string\Appendix~\chapterlabel} }}
\def\chapterreset{\gl@bal\advance\chapternumber by 1
   \ifnum\equanumber<0 \else\gl@bal\equanumber=0\fi
   \gl@bal\sectionnumber=0 \let\sectionlabel=\rel@x
   \ifcn@ \gl@bal\cn@@true {\pr@tect
       \xdef\chapterlabel{{\the\chapterstyle{\the\chapternumber}}}}%
    \else \gl@bal\cn@@false \gdef\chapterlabel{\rel@x}\fi }%MODIFICATION
%
%%%%%%%%%%%%%%%%%%%%%%%%%%%%%%%%%%%%%%%%%%%%%%%%%%%%%%%%%%%%%%%%%%%%%%%
%%
%%       Here is the draftmode feature
%%
%%       Use the following on the preliminary draft,
%%       puts time/date on each page in writes labels in margins
%%       and puts reference labels on the reference page.
%%       Putting \draft in the beginning of the paper causes it
%%       to be printed in draftmode.  use \nodraftlabels to get rid of
%%       eqn, ref, and fig labels in draft mode
%%
%%       Timestamp routine bug fixed October 30, 1991 by T.J. Allen
%%
%%%%%%%%%%%%%%%%%%%%%%%%%%%%%%%%%%%%%%%%%%%%%%%%%%%%%%%%%%%%%%%%%%%%%%%
%
\newif\ifdr@ftmode
\newtoks\r@flabeltoks
\def\draftmode{
   \tenpoint
   \baselineskip=24pt plus 2pt minus 2pt
   \dr@ftmodetrue
   \message{ DRAFTMODE }
   \writedraftlabels
   \def\timestring{\begingroup
     \count0 = \time \divide\count0 by 60
     \count2 = \count0  % the hour
     \count4 = \time \multiply\count0 by 60
     \advance\count4 by -\count0   % the minute
     \ifnum\count4<10 \toks1={0} % get a leading zero.
     \else \toks1 = {}
     \fi
     \ifnum\count2<12 \toks0={a.m.} %
          \ifnum\count2<1 \count2=12 \fi% Make midnight `12'
     \else            \toks0={p.m.} %
           \ifnum\count2=12 % keep noon `12'
           \else
           \advance\count2 by -12 % keep afternoon times < 12
           \fi
     \fi
%    \ifnum\count2=0 \count2 = 12\fi % make midnight `12'. %  There seems to
%    be a bug in TeX when checking a count which has the value 0.
     \number\count2:\the\toks1 \number\count4\thinspace \the\toks0
   \endgroup}%
   \def\draftdate{{{\tt preliminary version:}\space{\rm
                                  \timestring\quad\the\date}}}
\def\R@FWRITE##1{\ifreferenceopen \else \gl@bal\referenceopentrue
     \immediate\openout\referencewrite=\jobname.refs
     \toks@={\begingroup \refoutspecials \catcode`\^^M=10 }%
     \immediate\write\referencewrite{\the\toks@}\fi
     \immediate\write\referencewrite%
     {\noexpand\refitem{\the\r@flabeltoks[\the\referencecount]}}%
     \p@rse@ndwrite \referencewrite ##1}
\def\refitem##1{\r@fitem{##1}}
\def\REF##1##2{\reflabel##1 \REFNUM ##1\REFWRITE{\ignorespaces ##2}}
\def\Ref##1##2{\reflabel##1 \Refnum ##1\REFWRITE{ ##2}}
\def\REFS##1##2{\reflabel##1 \REFNUM ##1%
\gl@bal\lastrefsbegincount=\referencecount\REFWRITE{ ##2}}
\def\refs{\REFS\?}
\def\refc{\REF\?}
\let\refscon=\refc       \let\REFSCON=\REF
}
\def\nodraftlabels{\def\leqlabel##1{}\def\eqlabel##1{}\def\reflabel##1{}%
\def\leqlabel##1{}}
\def\writedraftlabels{
  \def\eqlabel##1{{\escapechar-1\rlap{\sevenrm\hskip.05in\string##1}}}%
  \def\leqlabel##1{{\escapechar-1\llap{\sevenrm\string##1\hskip.05in}}}%
  \def\reflabel##1{\r@flabeltoks={{\escapechar-1\sevenrm\string##1\hskip.06in%
}}}}

\nodraftlabels   % Make the default mode no labels
\dr@ftmodefalse  % Turn off draftmode
%
%%%%%%%%%%%%%%%%%%%%%%%%%%%%%%%%%%%%%%%%%%%%%%%%%%%%%%%%%%%%%%%%%%%%%%%%
%%
%%   Here come macros for equation numbering.
%%   (Equation numbers are modified in draft mode)
%%
%%   Sections are automatically numbered independently, unless
%%   one puts the command \sequentialequations
%%
%%%%%%%%%%%%%%%%%%%%%%%%%%%%%%%%%%%%%%%%%%%%%%%%%%%%%%%%%%%%%%%%%%%%%%%%
%
%

\def\eqn#1{\eqno\eqname{#1}\eqlabel#1}
 %MODIFICATION
%
%
\def\eqinsert#1{\noalign{\dimen@=\prevdepth \nointerlineskip
   \setbox0=\hbox to\displaywidth{\hfil #1}
   \vbox to 0pt{\kern 0.5\baselineskip\hbox{$\!\box0\!$}\vss}
   \prevdepth=\dimen@}}  %MODIFICATION

%

 %MODIFICATION

%
 %MODIFICATION
%
%\def\sequentialequations{\equanumber=-1}
%\def\sequentialequations{\rel@x \if\equanumber<0 \else
%  \gl@bal\equanumber=-\equanumber \gl@bal\advance\equanumber by -1 \fi }
%                         %MODIFICATION
%
%%%%%%%%%%%%%%%%%%%%%%%%%%%%%%%%%%%%%%%%%%%%%%%%%%%%%%%%%%%%%%%%%%%%%%%%
%%
%%  Here come reference macros  (Modified for the new version of phyzzx)
%%
%%%%%%%%%%%%%%%%%%%%%%%%%%%%%%%%%%%%%%%%%%%%%%%%%%%%%%%%%%%%%%%%%%%%%%%%
%
%
\def\refout{\par\penalty-400\vskip\chapterskip
   \spacecheck\referenceminspace
   \ifreferenceopen \Closeout\referencewrite \referenceopenfalse \fi
   \line{\ifpr@printstyle\twelverm\else\fourteenrm\fi
         \hfil REFERENCES\hfil}\vskip\headskip
   \input \jobname.refs
   }
\def\ACK{\par\penalty-100\medskip \spacecheck\sectionminspace
   \line{\ifpr@printstyle\twelverm\else\fourteenrm\fi
      \hfil ACKNOWLEDGEMENTS\hfil}\nobreak\vskip\headskip }
\def\tabout{\par\penalty-400
   \vskip\chapterskip\spacecheck\referenceminspace
   \iftableopen \Closeout\tablewrite \tableopenfalse \fi
   \line{\ifpr@printstyle\twelverm\else\fourteenrm\fi\hfil TABLE CAPTIONS\hfil}
   \vskip\headskip
   \input \jobname.tabs
   }
\def\figout{\par\penalty-400
   \vskip\chapterskip\spacecheck\referenceminspace
   \iffigureopen \Closeout\figurewrite \figureopenfalse \fi
   \line{\ifpr@printstyle\twelverm\else\fourteenrm\fi\hfil FIGURE
CAPTIONS\hfil}
   \vskip\headskip
   \input \jobname.figs
   }
\def\masterreset{\begingroup\hsize=\pagetextwidth
   \global\pagenumber=1 \global\chapternumber=0
   \global\equanumber=0 \global\sectionnumber=0
   \global\referencecount=0 \global\figurecount=0 \global\tablecount=0
   \endgroup}
%

% % % % % % % % % % % % % % % % % % % % % % % % % % % % % % % % % % % %
%%%%%%%%%%%%%%%%%%%%%%%%%%%%%%%%%%%%%%%%%%%%%%%%%%%%%%%%%%%%%%%%%%%%%%%%%
%%
%%      Various little user definitions
%%
%%%%%%%%%%%%%%%%%%%%%%%%%%%%%%%%%%%%%%%%%%%%%%%%%%%%%%%%%%%%%%%%%%%%%%%
%

\def\12{{1\over2}}

\def\sla{\raise.15ex\hbox{$/$}\kern-.57em}
\def\leaderfill{\leaders\hbox to 1em{\hss.\hss}\hfill}
\def\dual{{\,^*\kern-.20em}}
\def\bx{{\vcenter{\hrule height 0.4pt
      \hbox{\vrule width 0.4pt height 10pt \kern 10pt
        \vrule width 0.4pt}
      \hrule height 0.4pt}}}
\def\inner{\,{\vcenter{
      \hbox{ \kern 4pt
        \vrule width 0.5pt height 7pt}
      \hrule height 0.5pt}}\,}
\def\sqr#1#2{{\vcenter{\hrule height.#2pt
      \hbox{\vrule width.#2pt height#1pt \kern#1pt
        \vrule width.#2pt}
      \hrule height.#2pt}}}
\def\rect#1#2#3#4{{\vcenter{\hrule height#3pt
      \hbox{\vrule width#4pt height#1pt \kern#1pt
        \vrule width#4pt}
      \hrule height#3pt}}}

\def\bx{{\vcenter{\hrule height 0.4pt
      \hbox{\vrule width 0.4pt height 10pt \kern 10pt
        \vrule width 0.4pt}
      \hrule height 0.4pt}}}

\def\up#1{\leavevmode \raise.16ex\hbox{#1}}
\def\twiddle{\lower.9ex\rlap{$\kern-.1em\scriptstyle\sim$}}
\def\bigtwiddle{\lower1.ex\rlap{$\sim$}}
\def\gtwid{\mathrel{\raise.3ex\hbox{$>$\kern-.75em\lower1ex\hbox{$\sim$}}}}
\def\ltwid{\mathrel{\raise.3ex\hbox{$<$\kern-.75em\lower1ex\hbox{$\sim$}}}}
\def\square{\kern1pt\vbox{\hrule height 1.2pt\hbox{\vrule width 1.2pt\hskip 3pt
   \vbox{\vskip 6pt}\hskip 3pt\vrule width 0.6pt}\hrule height 0.6pt}\kern1pt}
\def\tdot#1{\mathord{\mathop{#1}\limits^{\kern2pt\ldots}}}

\def\pmb#1{\setbox0=\hbox{#1}    %  POOR MAN'S BOLD
  \kern-.025em\copy0\kern-\wd0
  \kern  .05em\copy0\kern-\wd0
  \kern-.025em\raise.0433em\box0 }

    %%DALEMBERTIAN, USED TO BE \box

\def\prl{\journal Phys. Rev. Lett. }

\hyphenation{anom-aly anom-alies coun-ter-term coun-ter-terms}
\def\inv{^{\raise.15ex\hbox{${\scriptscriptstyle -}$}\kern-.05em 1}}

\def\Dsl{\,\raise.15ex\hbox{/}\mkern-13.5mu D} %this one can be subscripted
\def\dsl{\raise.15ex\hbox{/}\kern-.57em\partial}

       %pound sterling
\def\boxeqn#1{\vcenter{\vbox{\hrule\hbox{\vrule\kern3pt\vbox{\kern3pt
        \hbox{${\displaystyle #1}$}\kern3pt}\kern3pt\vrule}\hrule}}}
\def\mbox#1#2{\vcenter{\hrule \hbox{\vrule height#2in
                \kern#1in \vrule} \hrule}}  %e.g. \mbox{.1}{.1}
%
%%       matters of taste
%%  \def\tilde{\widetilde} \def\bar{\overline} \def\hat{\widehat}
%%

\def\psibar{\overline\psi}

\def\darr#1{\raise1.5ex\hbox{$\leftrightarrow$}\mkern-16.5mu #1}
 %pound sterling
\def\roughly#1{\raise.3ex\hbox{$#1$\kern-.75em\lower1ex\hbox{$\sim$}}}
%
%
%%%%%%%%%%%%%%%%%%%%%%%%%%%%%%%%%%%%%%%%%%%%%%%%%%%%%%%%%%%%%%%%%%%%%%%%
%%
%%   Miscellaneous macros
%%
%%%%%%%%%%%%%%%%%%%%%%%%%%%%%%%%%%%%%%%%%%%%%%%%%%%%%%%%%%%%%%%%%%%%%%%%
%
\def\ack{\ACK}   % make new phyzzx compatible with old phyzzx
\font\titlerm=cmr10 scaled \magstep 4
\def\TITLEPAGE{\frontpagetrue\pageno=1\pagenumber=1}

\def\WISCONSIN{\vskip15pt\vbox{\hbox{\centerline{\it Department of Physics}}
        \vskip 0pt
  \hbox{\centerline{\it University of Wisconsin, Madison, WI 53706 USA}}}}

\def\TITLE#1{\vskip 1in \centerline{\titlerm #1}}
\def\MORETITLE#1{\vskip 19pt \centerline{\titlerm #1}}
\def\AUTHOR#1{\vskip .5in \centerline{#1}}

\def\ABSTRACT#1{\vskip .5in \vfil \centerline{\twelvepoint \bf Abstract}
        #1 \vfil}
\def\ENDTITLEPAGE{\vfill\eject\pageno=2\pagenumber=2}%\hsize=\pagetextwidth}
\let\letterhead=\MADPHHEAD

\def\underwig#1{{
\setbox0=\hbox{$#1$}
\setbox1=\hbox{}
\wd1=\wd0
\ht1=\ht0
\dp1=\dp0
\setbox2=\hbox{$\rm\widetilde{\box1}$}
\dimen@=\ht2 \advance \dimen@ by \dp2 \advance \dimen@ by 1.5pt
\ht2=0pt \dp2=0pt
\hbox to 0pt{$#1$\hss} \lower\dimen@\box2
}}
\def\bunderwig#1{{
\setbox0=\hbox{$#1$}
\setbox1=\hbox{}
\wd1=\wd0
\ht1=\ht0
\dp1=\dp0
\setbox2=\hbox{$\seventeenrm\widetilde{\box1}$}
\dimen@=\the\ht2 \advance \dimen@ by \the\dp2 \advance \dimen@ by 1.5pt
\ht2=0pt \dp2=0pt
\hbox to 0pt{$#1$\hss} \lower\dimen@\box2
}}
\def\journal#1&#2(#3){\unskip, \sl #1~\bf #2 \rm (19#3) }
                    % Journal reference. Alignment
                    % tabs & set off name, vol, year, page
\def\npjournal#1&#2&#3&#4&{\unskip, #1~\rm #2 \rm (#3) #4}
\gdef\prjournal#1&#2&#3&#4&{\unskip, #1~\bf #2, \rm #4 (#3)}

\let\int=\intop         
\catcode`@=12 % at signs are no longer letters
\masterreset
%
%%%%%%%%%%%%%%%%%%% End enhanced phyzzx macro %%%%%%%%%%%%%%%%%%%%%%%%%%%%%%%
%%%%%%%%%%%%%%%%%% Beginning of paper proper %%%%%%%%%%%%%%%%%%%%%%%%%%%%%%%%
%\draft
%\preprintstyle
%\twocolumn
\normalspace    % use spacing and a half
%\doublespace
\chapterheadstyle={\bf}     % this is for making chapterheadings boldfaced
\overfullrule=0pt  % this eliminates the annoying blackboxes on overfull lines

\def\prl{{\it Phys. Rev. Lett.\/ }}
\def\mpl{{\it Mod. Phys. Lett.\/ }}

\def\psibar{\overline{\psi}}

\let\SS=\ss
\TITLEPAGE
\rightline{{\tenpoint\baselineskip=12pt
           \vtop{\hbox{\strut MAD/TH-92-03}
                 \hbox{\strut hep-th/9207093}
                 \hbox{\strut June 1992}}}}
%\rightline{\it Revised}
\TITLE{ Collective Coordinate Action for Charged }
\MORETITLE{ Sigma-Model Vortices in Finite Geometries }
\AUTHOR{ Theodore J. Allen{${}^*$} }
\footnote{*}{tjallen@wishep.physics.wisc.edu}
\WISCONSIN

\ABSTRACT{ In this Letter the method of Lund is applied to formulate a
variational principle for the motion of charged vortices in an effective
non-linear Schr\"{o}dinger field theory describing finite size two-dimensional
quantum Hall samples under the influence of an arbitrary perpendicular magnetic
field.  Freezing out variations in the modulus of the effective field yields a
$U(1)$ sigma-model.  A duality transformation on the sigma-model reduces the
problem to finding the Green function for a related electrostatics problem.
This duality illuminates the plasma analogy to the Laughlin wave function. }

\ENDTITLEPAGE
%\begindoublecolumns

The fractional quantum Hall effect shares a number of remarkable features with
the macroscopic quantum phenomena of superconductivity and superfluidity.   It
occurs only at very low temperatures, is dissipationless and is a result of a
broken (translational) symmetry.\Ref\GiddWil{S.B.~Giddings and F.~Wilczek, \mpl
{\bf A5} (1990) 635.}  It is reasonable to assume that there should exist an
effective field theory\Ref\GLIdea{S.~Girvin and A.~MacDonald, \prl {\bf 58}
(1987) 1252\semi S.M.~Girvin, {\it in }  {\sl The Quantum Hall Effect},
R.E.~Prange and S.M.~Girvin, eds., (Springer Verlag, New York,  1987) p.\
389\semi S.C.~Zhang, T.H.~Hansson and S.~Kivelson, \prl {\bf 62}  (1989)
82\semi N.~Read, \prl {\bf 62} (1989) 86.} for the macroscopic quantum state of
fractional quantum Hall systems.  As in the effective theories of the other
macroscopic quantum phenomena, there are vortex excitations in the effective
theory of the quantum Hall effect.  However, unlike their better-known cousins,
these vortices are point-like (because of the two-dimensional nature of the
effect) and play the central role in the physics of this phenomenon.

An effective field theory, or Ginzburg-Landau, description of the fractional
quantum Hall effect provides a beautiful characterization of the fractionally
charged quasiparticles necessary for fractional conductivity as charged vortex
excitations of the quantum Hall fluid.   The quantum mechanics of a charged
vortex yields the lowest Landau level states of a particle of equal
charge,\Ref\AllBord{T.J.~Allen and A.J.~Bordner, {\it Charged Vortex Dynamics
in Ginzburg-Landau Theory of the Fractional Quantum Hall Effect}, UW-Madison
preprint MAD/TH-92-02, hep-th/9206073.} even though vortex mechanics is quite
different in detail from the usual charged particle mechanics.  Because
vortices are extended field configurations, the geometry of the sample in which
they occur determines the Hamiltonian and hence their dynamics.  The purpose of
this Letter is to derive the action for arbitrary sample geometry.

Vortices in an infinite plane have a pair-wise logarithmic interaction as well
as a single-vortex interaction with the background magnetic field.  When
boundaries are present, there is an additional vortex---boundary interaction
which can be written with the help of image vortices in simple geometries.

The logarithmic interaction is analogous to vortex-vortex interactions in fluid
mechanics.  The center of one vortex is carried along with the bulk motion of
fluid around another.  The external vector potential corresponds to a
background motion of the fluid with vorticity density equal to the applied
magnetic field.  Motion in a uniform field, for instance, is analogous to a
uniform rotation about some point.  In this Letter, the duality between this
viewpoint and an electrostatic description will be used to derive the
expression for the collective Hamiltonian in a two-dimensional bounded region
for an arbitrary applied magnetic field.

The derivation of the collective action for the motion of the vortex centers
starts with the elegant idea of F.~Lund,\Ref\Lund{F. Lund, \it Phys. Lett. \bf
A 159 \rm (1991) 245.} which is to substitute the collective vortex ansatz for
the field configuration with  vortices at ${\bf X}^{\SS A}$, ${\SS A} = 1,
\ldots, N$,  into the effective field theory action and reduce the action to a
function of the collective coordinates,
$$S_V[\{{\bf X}^{\SS A}(t)\}] = S_{\caps eft}[\psi({\bf x},t;\{{\bf X}^{\SS
A}(t)\})].\eqn\CollAction$$
Here we follow Lund's method but we must be careful in choosing the
multi-vortex ansatz, $\psi=\psi({\bf x},t;\{{\bf X}^{\SS A}(t)\})$.  The usual
ansatz in an infinite  region,
$$\psi=\sqrt{\rho_0}\exp\left\{i\sum_{\SS A} n_{\SS A} \tan^{-1}\left({x_2 -
X^{\SS A}_2(t)\over x_1-X^{\SS A}_1(t)}\right)\right\},\eqn\multivorI$$
with $\rho_0$ constant  is not valid because it violates the boundary
conditions.

The correct boundary conditions on the ansatz follow from considering
variations on the field $\psi$ which are arbitrary at the boundary.  Thus,
along with the usual Euler-Lagrange equations of motion in the region $\Sigma$,
one must impose the Ginzburg-Landau conditions
$$\hat{\bf n}\cdot{\partial L\over\partial(\nabla\psibar)}
\bigg|_{\partial\Sigma}=0,\eqn\GLbdry$$
at a free boundary.  In a fluid mechanical system this would enforce the
requirement that the normal velocity of the fluid vanish at the walls of a
container.  In a quantum mechanical system the boundary conditions force the
vanishing of the normal electric current.

In Ref.\ [\AllBord] the non-linear Schr\"{o}dinger Lagrangian,
$${\cal L}=i\psibar\dot\psi - {\alpha\over|\psi|^2}|(\nabla - ie{\bf A})
\psi|^2 - V(|\psi|^2),\eqn\Lag$$
for a charged field $\psi$  in a background magnetic field, $B({\bf x}) =
\nabla\times{\bf A}$, was proposed as an effective theory for the fractional
quantum Hall effect  which has charged vortex solutions.  The equations of
motion for this Lagrangian are solved by configurations of the effective field
$$\psi = \sqrt{\rho_0}\exp (i\phi),\eqn\Solns$$
with constant $\rho_0$, and harmonic phase,
$$ \nabla^2\phi = 0, \eqn\EoMGauge $$
in a Coulomb gauge, $\nabla\cdot{\bf A} = 0$.

Because \EoMGauge\ is linear, single vortex excitations can be superposed to
form multi-vortex excitations, $\phi = \sum_{\SS A}n_{\SS A}\phi_{\SS A}$.
The $\phi_{\SS A}$ are unit vortex solutions and the integers $n_{\SS A}$ are
the strengths of the vortices centered at ${\bf X}^{\SS A}$. The condition
characterizing a unit vortex with center ${\bf X}^{\SS A}$ is
$$\nabla\times\nabla\phi_{\SS A}({\bf x}) = 2\pi\delta^{(2)}({\bf x} - {\bf
X}^{\SS A}).\eqn\vort$$
Eqs.\ \EoMGauge\ and \vort\ are much like the Bianchi identity and equations of
motion for a field strength $\nabla\phi$.  It is well known that one can
dualize such equations by using a volume form.  In this case we shall use the
area form $\varepsilon = {{\scriptscriptstyle\phantom{-} 0}\,\,\,
{\scriptscriptstyle 1}\,\choose{\scriptscriptstyle -1}\,\,\,
{\scriptscriptstyle 0}\,}$ to  define a dual potential $\chi$,
$$\nabla\phi_{\SS A} = \varepsilon\cdot\nabla\chi_{\SS A}.\eqn\dualCHI$$
The equations \EoMGauge\ and \vort\ for $\phi$ become
$$\eqalign{\nabla^2\chi_{\SS A}({\bf x}) &= -2\pi\delta^{(2)}({\bf x} - {\bf
X}^{\SS A}),\cr \nabla\times\nabla\chi_{\SS A} &= 0.\cr}\eqn\dualEoM$$
It is useful to dualize the gauge and field-strength equations for $\bf A$ as
well,
$$\eqalign{{\bf A} &= \varepsilon\cdot\nabla\Upsilon,\cr
\nabla\times\nabla\Upsilon &=0,\cr \nabla^2\Upsilon &= -B({\bf
x}).\cr}\eqn\Adual$$
The boundary conditions on $\phi$ in eq.\ \Solns\ following from the Lagrangian
\Lag,
$$\hat{\bf n}\cdot\left(\nabla\phi - e{\bf A}\right)\big|_{\partial\Sigma} = 0,
\eqn\bdryPHI$$
in the dual variables become
$$\hat{\bf n}\times\left(\nabla\chi_{\SS A} -
e\nabla\Upsilon\right)=0.\eqn\bdryCHI$$
These conditions can generally be taken to be the vanishing of both $\chi_{\SS
A}$ and  $\Upsilon$ on smooth boundaries because $\hat{\bf n}\times\nabla$ is
the derivative along the boundary.  The $\chi_{\SS A}$ for several vortices can
then be superposed without changing the boundary conditions.

The dual variables are useful for constructing the collective action for the
charged vortices.   According to the prescription \CollAction, we  substitute
the ansatz \Solns\ in the form of
$$\psi=\sqrt{\rho_0}\exp\left(i\sum_{\SS A} n_{\SS A}\phi_{\SS A}({\bf x},{\bf
X}^{\SS A}(t))\right),\eqn\multivorII$$
into the action \Lag.  In eq.\ \multivorII\ each $\phi_{\SS A}$ satisfies
\vort, $\rho_0$ is taken to be constant, and the total phase obeys boundary
conditions \bdryPHI. The result is the $U(1)$ sigma-model action
$$ L = \int d^2\! x\left(-\sum_{\SS A}\rho_0n_{\SS A}\dot\phi_{\SS A} -
\alpha\sum_{{\SS A},{\SS B}}(n_{\SS A}\nabla\phi_{\SS A} - e{\bf
A})\cdot(n_{\SS B}\nabla\phi_{\SS B} - e{\bf A})\right).\eqn\LagPHI$$
Following Lund\refmark{\Lund} we put a fictitious extra boundary in the sample
consisting of a small circle around each vortex and cuts ${\cal S}_{\SS A}$
from each vortex point ${\bf X}^{\SS A}$ to a fixed point ${\bf x}_0$ on the
boundary.  This has the  effect of making every quantity in the integrand
single-valued in the  cut region.  Each $\phi_{\SS A}$ jumps by $-2\pi$ on its
cut ${\cal S}_{\SS A}$. The action following from \LagPHI\ reduces to
$$\eqalign{S_V = \int dt\biggl[&\sum_{\SS A}\pi\rho_0n_{\SS
A}\epsilon_{ab}X^{\SS A}_a \dot X^{\SS A}_b - 4\pi\alpha\sum_{{\SS A}<{ \SS
B}}n_{\SS A}n_{\SS B}\int_{{\cal S}_{\SS
A}}d\pmb{$\ell$}\cdot\varepsilon\cdot\nabla\phi_{\SS B}\cr & -
2\pi\alpha\sum_{\SS A} n_{\SS A}^2\int_{{\cal S}_{\SS A}}
d\pmb{$\ell$}\cdot\varepsilon\cdot\nabla\phi_{\SS A} + 4\pi\alpha e\sum_{\SS
B}n_{\SS B}\int_{{\cal S}_{\SS B}}d\pmb{$\ell$}\cdot\varepsilon\cdot{\bf
A}\biggr],\cr} \eqn\SphiI$$
where terms not depending on the vortex centers have been dropped and the cuts
${\cal S}_{\SS A}$ from the vortex ${\SS A}$ to a fixed point on the boundary
are explicitly
$${\cal S}_{\SS A}=\big\{{\bf x}(s,t)\big|\,\,0\leq s\leq 1,\quad {\bf x}(0,t)=
{\bf X}^{\SS A}(t), \quad {\bf x}(1,t) = {\bf x}_0\in\partial\Sigma\big\}.
\eqn\SA$$
To obtain \SphiI\ we used the divergence theorem and the fact that the boundary
conditions \GLbdry\ ensure that there is no contribution from the physical
boundary to the vortex action $S_V$.  We also dropped as negligible  the
contributions from the small boundary circles around each vortex.  The first
term results from a double integration by parts.\refmark{\Lund}   In the
dualized variables the collective action becomes
$$\eqalign{S_V[\{{\bf X}^{\SS A}\}] = \int dt\biggl\{&\sum_{\SS
A}\pi\rho_0n_{\SS A}\epsilon_{ab}X^{\SS A}_a\dot X^{\SS A}_b +
4\pi\alpha\sum_{{\SS A}<{\SS B}}n_{\SS A}n_{\SS B}\int_{{\cal S}_{\SS
A}}d\pmb{$\ell$}\cdot\nabla\chi_{\SS B} \cr  &+2\pi\alpha\sum_{\SS A}n^2_{\SS
A}\int_{{\cal S}_{\SS A}}  d\pmb{$\ell$}\cdot\nabla\chi_{\SS A} - 4\pi\alpha
e\sum_{\SS B}n_{\SS B}\int_{{\cal S}_{\SS
B}}d\pmb{$\ell$}\cdot\nabla\Upsilon\biggr\}.\cr}\eqn\SphiII$$
We observe that $\chi_{\SS B}({\bf x})$ is the Green function $G_D({\bf x},{\bf
X}^{\SS B})$ satisfying zero Dirichlet boundary conditions. Because $\Upsilon$
satisfies $\nabla^2\Upsilon = -B({\bf x})$, $\Upsilon|_{\partial\Sigma} = 0$,
we also have $\Upsilon({\bf x})={1\over2\pi}\int_\Sigma G_D({\bf x},{\bf
x^\prime}) B({\bf x^\prime})\,d^2\! x^\prime$.  The expression for the charged
vortex  collective action in a finite size region  under the influence of an
arbitrary external magnetic field can be written in terms of the Dirichlet
Green  function for the region as
$$\eqalign{ S_V[\{{\bf X}^{\SS A}\}] = \int dt\biggl\{&\sum_{\SS
A}\pi\rho_0n_{\SS A}\epsilon_{ab}X^{\SS A}_a\dot X^{\SS A}_b -
4\pi\alpha\sum_{{\SS A}<{\SS B}}n_{\SS A}n_{\SS B} G_D({\bf X}^{\SS A},{\bf
X}^{\SS B})\cr &-2\pi\alpha\sum_{\SS A}n^2_{\SS A}G_D^{\rm reg}({\bf X}^{\SS
A},{\bf X}^{\SS A})\cr &+ 2\alpha e\sum_{\SS A}n_{\SS A}\int_\Sigma d^2\!{x}\,
G_D({\bf X}^{\SS A},{\bf x})B({\bf x})\biggr\}.\cr} \eqn\SphiIII$$
Here we have ignored an infinite self-energy contribution to $S_V$ which does
not depend upon the vortex centers by using the regularized Green function
$$ G_D^{\rm reg}({\bf x},{\bf x}) = \lim_{{\bf y}\to{\bf x}}\left(G_D({\bf
x},{\bf y}) + \ln|{\bf x} - {\bf y}|\right).\eqn\regGreenFcn$$

The vortex collective Hamiltonian is thus the same as the interaction energy
for an assembly of charges, $q^{\SS A} = n_{\SS A}$, in the bounded region
$\Sigma$ interacting with a background charge density proportional to the
applied magnetic field, $B({\bf x})/2\pi$.  The first term in the action,
however, is not the usual kinetic term, so the motion of the charged vortices
is quite different from that of point charges.  In particular, the vortices do
not  experience an acceleration which is proportional to the gradient of the
potential, but rather follow the instantaneous equipotentials described by the
last three terms of \SphiIII\ with speeds proportional to the gradient of the
potential.  A single vortex has as its orbits the equipotentials of a region
$\Sigma$ with charge density $B({\bf x})/2\pi$, whose  boundary,
$\partial\Sigma$, is held at constant potential.  The fact that the phase space
of the vortices is just the two-dimensional sample itself together with the
identification of the vortex Hamiltonian as an electrostatic energy
illuminates the plasma analogy of the quantum Hall effect.  If the density
matrix computed from the Laughlin wave function is identified with the density
matrix of a canonical ensemble of vortices, we have immediately
$$ |\Psi_{\rm Laughlin}|^2 = \varrho_{\rm Laughlin} = \varrho_{V} = \exp(-\beta
H_V).\eqn\VortexEnsemble$$
%%
%

%\enddoublecolumns

\ack

It is a pleasure to thank L.~Durand for a useful conversation which motivated
this investigation, and R.P.~Springer for helpful comments on the manuscript.
This work was supported in part by DOE grant No.\ DE-AC02-76-ER00881.

\refout
\bye